# Concurrent Development of Model and Implementation

**Andrew M. Gravell, Yvonne Howard, Juan C. Augusto, Carla Ferreira, and Stefan Gruner**
University of Southampton, Southampton SO17 1BJ, England
Phone 0(+44)23 8059 2741, Fax 0(+44) 8059 3045
Email amg@ecs.soton.ac.uk, ymh@ecs.soton.ac.uk, etc.

*Abstract*

This paper considers how a formal mathematically-based model can be used in support of evolutionary software development, and in particular how such a model can be kept consistent with the implementation as it changes to meet new requirements. A number of techniques are listed can make use of such a model to enhance the development process, and also ways to keep model and implementation consistent. The effectiveness of these techniques is investigated through two case studies concerning the development of small e-business applications, a travel agent and a mortgage broker. Some successes are reported, notably in the use of rapid throwaway modelling to investigate design alternatives, and also in the use of close team working and model-based trace-checking to maintain synchronisation between model and implementation throughout the development. The main areas of weakness were seen to derive from deficiencies in tool support. Recommendations are therefore made for future improvements to tools supporting formal models which would, in principle, make this co-evolutionary approach attractive to industrial software developers. It is claimed that in fact tools already exist that provide the desired facilities, but these are not necessarily production-quality, and do not all support the same notations, and hence cannot be used together.

# 1. *Introduction*

Over the last two decades, traditional top-down approaches to software engineering, such as the Waterfall model [19], have gradually lost their dominance in favour of evolutionary and iterative methods. In an evolutionary approach, such as the Unified Process [20], previously separate phases such as requirements analysis, design and implementation are allowed to overlap, and the requirements themselves can change or evolve while the system is being developed. In an iterative approach, such as eXtreme Programming [4], the system is constructed through a series of versions, each of which delivers additional features, as agreed at the start of each iteration. Iterative approaches naturally support and encourage the evolution of requirements and design, since these typically change with each iteration; they are also therefore called agile methods [9]. These newer methods of software development were created in response to the failure, or perceived failure, of top-down approaches to deliver the systems that meet or beat their specification, deadline and budget. Indeed, the increasing use of the term "software development" rather than "software engineering" is indicative of the move to evolutionary approaches.

The currently accepted approach in the formal methods community is top-down. Designers should use a formal, mathematically-based notation such as B [1] or Z [22] to write a specification, which is then refined, gradually adding detail until a level has been reached that can straightforwardly or automatically be turned into program code. Associated with this process are proof obligations; these could be ignored, checked informally, or else proved correct, as for example in the B method.

This top-down formal stepwise development method has not been widely adopted
(except for safety-critical systems where additional regulatory requirements exist). Part of the reason for this failure, we speculate, is that programmers are forced to wait until designers have completed the specification. Of course, for a small system, the programmers will also be the designers – but larger teams typically have some specialisation of development roles. Completing a specification can be a time-consuming activity, since it requires designers to think through all special and error cases, and in fact these cases sometimes come to light only after implementation has started. In addition, in many cases there is not the time, nor the skills, required to discharge all proof obligations. Where proof is not being used, it is not immediately clear that writing formal specifications and designs (formal models) offers any significant benefits over less formal but more popular modelling notations such as the UML [21].

This paper considers two issues. Firstly how a formal model can be incorporated in an evolutionary development, and secondly what benefits there may be in having a formal model in situations where it is not necessary, or not desired, to carry out formal proofs. An example would be the implementation of a business application, which is unlikely to be safety-critical, though it might be mission-critical.

With regard to the first of these, in an evolutionary development, implementation work will begin with a specification that is at most partially complete – which is why it is more helpful to use the phrase "formal model" rather than "formal specification". During the development, both the implementation and model will evolve. The question arises, how can these two evolving artefacts be kept consistent? Indeed, what does consistency mean in this context?

With regard to the benefits, formal models are written in notations with defined semantics. This makes them amenable to automated analysis and other tool support, beyond simply checking syntax and type-correctness. In particular, there has been considerable use recently of model-checking in industry [8, 13], albeit mostly for hardware and protocol development. Another area of interest is model-based test-case generation [10]. A final area of tool support is animation or visualisation of the model.

By contrast, precise semantics for the UML remains a research issue, limiting the scope for automated UML model-checking or test-case generation.

The approach described in this paper is illustrated by development case studies where we have attempted to develop concurrently both a model, or models, and an implementation. Both case studies are e-business applications. The first of these was a travel agent system, allowing users to book hotels and reserve cars, and the second was a mortgage broker, allowing users to obtain quotes for mortgages and house insurance allowing them to buy a house. The second was a similar system, concerning a mortgage broker, who coordinates offers from a number of mortgage lenders and property insurance companies, and helps the would-be house buyer to choose the best combination. This application is noteworthy only in that the implementation involves both classical (ACID) transactions, and long-running transactions, that require message-queuing and compensation mechanisms to implement.

The case studies build on our previous work [2, 3, 11] where we have investigated the use of formal modelling in e-business software development, which led us to propose this co-evolutionary approach.

The remaining sections of the paper cover benefits of formal models, the application of the proposed techniques in the case studies, and an evaluation their effectiveness, related and future work.

## 2. *Benefits of Formal Models*

This section considers potential benefits that can be obtained from having a formal, mathematical model even in cases where it is not intended to prove (formally or informally) that the implementation conforms to the model.

One of the claimed benefits of formal methods is the extra insight gained from giving a precise definition of system behaviour [23]. Similar claims are also made with respect to producing a diagrammatic model using a notation such as the UML. Insight is hard to quantify, however, so it is difficult to measure these benefits, or to compare formal and diagrammatic approaches.

The clearest benefits of having a formal model are those that follow from the availability of tool support, as described below. The U2B tool [6], for example, can generate formal models from a diagrammatic UML model plus annotations, allowing the benefits of each approach to be combined.

### 2.1 Animation

Animation or visualisation of a model allows the developer to step through an execution sequence of the model from state to state. It requires the model to be written in a form that can be "executed", though the notion of execution here is not limited to imperative programming. The ProB tool [7], for example, uses constraint logic programming to determine the set of states that can follow the current one in a B model even if the operations have non-deterministic definitions. ProB supports a point and click user interface which provides a lightweight way of checking that a model defines the required behaviour. States can be displayed graphically or in textual form. Animation can be used to communicate the intention and consequences of a model in a more dynamic way than reading its mathematical definition. (Concrete examples help us to understand abstract definitions.) Clearly, animation can only be used to explore a limited number of paths through state transition graph, but it can help to eliminate simple errors before a more systematic, but more expensive, error removal process is applied. It can help to avoid misunderstandings that might occur with requirements expressed solely in natural language. By comparison, animation facilities for UML models seem relatively limited. (With executable UML [17], a platform dependent model can be produced and executed. This is not, as yet, a mainstream approach to using the UML.)

### 2.2 Model-Checking

Model-checking tools support exhaustive investigation of a system's state space, which typically this has to be finite. They can be used to detect states that do not satisfy an invariant, or to find execution paths that do not satisfy properties expressed as a temporal logic formula. Note that the emphasis, as with testing, is on finding errors, not proving their absence. This is because the model being checked is usually a simplified version of the real system: absence of errors in the model does not guarantee the system itself is correct. Conversely, errors found in the model usually indicate problems in the system itself, though not always, because the model may be over-simplified or simply wrong.

It is possible to create a model, or series of models, to explore the consequences of design decisions, for example whether to use synchronous or asynchronous message-passing. The model-checker can help the designer to determine what properties are and are not satisfied. This technique is valuable in technology exploration and risk-management, so it has been called "risk-driven" model-checking [3]; another appropriate name is "throwaway modelling" as there is no obligation to keep the model once it has served its purpose.

### 2.3 Assertions

Formal models, particularly those based on pre- and post-conditions, are naturally associated with the use of assertions in code. The AsmL [12] toolkit, for example, can automatically insert assertions at appropriate points in the code to check that each operation conforms to its specification. This approach is most helpful in debugging and error location, as assertion checking is usually disabled in production software.

### 2.4 Model-Based Testing

It is possible to use a formal model to assist in testing. Classic techniques such as equivalence partitioning and boundary testing can be applied to the definition of operations given in the model. Mathematical formulas are converted into disjunctive normal form to discover cases of interest, and constraint satisfaction to determine values that witness each such cases. This approach has been an area of active research for a number of years [10], and is now starting to deliver tools of industrial relevance [15].

Clearly, properties postulated during model-checking could be used to generate test cases, either systematically or automatically.

It is worth noting, however, that writing good test cases is essentially a creative activity. Automated test case generation can at most supplement, not replace, the use of skilled manual testing.

Another model-based approach to software testing involves the use of execution traces or event logs. It is common for production software to support event logging or tracing of some kind. Typically, traces can be rather large and it is appropriate to automatic analysis. It is possible to write special purpose trace analysis tools, for example using Perl or a similar scripting language. Model-based trace-checking [14] uses instead a model-checker as the analysis tool; a trace and a model are supplied and the model-checker determines whether or not the given trace conforms to the supplied model. An alternative name for the technique is trace-driven model-checking. ProB provides this facility directly. With a little ingenuity, it is possible to use other popular model-checkers such as Spin [13] in this fashion as well. Trace-checking has the advantage relative to assertion-checking that it can be run off-line and after the fact. Global invariants can also be checked more easily. Trace-checking only supplements other testing techniques – there has to be a test-case or user driving the system in order for there to be a trace to check. As with animation and model-checking, trace-checking is only applicable where the model is "executable", albeit that the requirement is weaker again, since here the input, output, pre- and post-operation values are all available – it is only necessary to test these, not to generate them.

### 2.5 Hybrid Approaches

The approaches listed above can be combined in various ways.

Model-checking, for example, can be seen as an extended or generalised kind of animation. The Spin tool supports user-driven as well as random, exhaustive and probabilistic state exploration. ProB also has both animation and model-checking facilities.

Large systems are usually divided into sub-components. It is possible to provide a single top-level model, and also a refined or decomposed model, which specifies each component and how they interact. Animation, model-checking, and model-based testing can then be applied at multiple levels. In cases where no model exists for a sub-component it may be possible to use the sub-component itself in, for example, model-checking.

### 2.6 Co-Evolution

In order to realise many of the above benefits, and in particular to use those techniques related to testing, it is necessary maintain close links with the model and the implementation. There is no point testing that a system conforms to a model which has been outdated by evolving requirements. It is therefore of interest to consider ways of maintaining synchronisation between model and specification. The classical approach is to verify the specification against the model using formal proof. This is rather heavyweight, and not appropriate in the context of an evolutionary or agile method. Alternative techniques for maintaining the consistency of model and implementation are:
- formal inspections of code against model
- close team work, with all developers in the same open plan office
- perhaps even pair "programming", where the modeller sits beside the programmer as they code, and vice versa

test-driven development [5], but using model-based testing instead of, or to supplement, manual or manually-written tests

In our case studies we have employed these techniques in order to obtain at least a subjective impression of their effectiveness.

## 3. *Case Studies*

To explore these possibilities for formal modelling, we have undertaken two case studies, both in the area of e-business applications. The first of these was a travel agent system, allowing users to book hotels and reserve cars, and the second was a mortgage broker, allowing users to obtain quotes for mortgages and house insurance allowing them to buy a house. The systems were implemented using Java servlets and JDBC to access a back-end relational database.

In these case studies there is a small development team of four, with a mix of skills. One of the developers is skilled in B/AMN modelling, the other three have a reading knowledge of B [1]. One of the developers is skilled in Spin/Promela [13] modelling. Three of the developers are familiar with Java server development. In such circumstances, it is natural for one or two of the team to work on modelling while the others prototype higher risk components, and then implement the full system. To support the risk-driven prototyping, we have modelled those aspects of the system first. Partly because of the existing skill mix, and partly because the different notations have different strengths, we have written models in both B/AMN and Spin/Promela model (to explore the state and the communication between distributed components). The models have been type-checked and validated using the simulation/animation facilities of standard tools supporting these formal notations. The B models were developed by first creating and annotating UML class diagrams, which were then converted to B machines using the U2B tool [6].

### 3.1 Formal Review and Close Team Working

To address the issue of consistency, we have used formal review meetings to compare the B/AMN model and implementation of the travel agency. These meetings have proved useful in finding problems both in the model, and also in the implementation. One difficulty [14] was that even after a week or two names for operations and classes could diverge. Though inspections are effective in finding such differences, it is tedious (and potentially error-prone) to rectify them.

In the mortgage broker case study, the team was more closely knit, as is usual in eXtreme Programming [4] and other agile methods. Instead of formal meetings, we relied instead on a close dialog between modellers and implementers, who were working in the same open plan office. In addition, ProB [7] and Promela animation were used to present the models to the developers. The dialog between modellers and developers meant that the top-level structure of model and implementation naturally have a close correspondence, together with common names for operations and classes. This is important, because it facilitate the use of model-based trace-checking.

In the first case study, we attempted to model the system at two levels, both from at the top-level, from the user's point of view, and also at the level of each component in our multi-tier implementation. Multi-level modelling proved time-consuming, and seemed to offer few benefits, in particular it was not well supported by the versions of the tools that were available at the time. Indeed, the only means we had of checking our lower level models was via formally proof of refinement. Though we had some success with these, it was very time-consuming. It is quite possible that the effort diverted to the proof could have been more effectively employed in other ways.

### 3.2 Rapid Modelling

The distinguishing feature of the second case study is that mortgage lenders and insurance companies respond asynchronously to these requests, allowing time for off-line credit checking, say. To track the progress of their request for quotation, the user must query the mortgage broker periodically to discover what offers they have received so far. The user can then accept one of these offers, in which case the broker will contact that mortgage lender and insurance company to commit them to their initial quotations, and inform the other lenders and insurers who quoted that they have been unsuccessful.

A potential issue with the mortgage broker system arises from the asynchronous nature of communications between the various parties. We wanted to explore the possible effects of messages being delayed and even lost. We decided to use rapid modelling to address this issue. A drastically simplified model of the system was created, with only one user, and minimal state for each component of the system.
We also modelled the messages passing between parties as a set. The next message to be delivered was selected using a non-deterministic choice; alternatively the message could be randomly dropped or deleted instead of being delivered. This simulates the behaviour of an asynchronous and unreliable network [16].

The question we were interested in understanding is whether the Broker believes they have committed a deal, but the selected lender doesn't, or vice versa. Using Spin, which is based on temporal logic, this property can naturally be expressed by a formula such as "whenever the broker requests a lender to commit an offer, eventually that lender does commit". The ProB system, however, is particularly suited to exploration of invariants. Fortunately we were able to capture the desired property in a different way, with a global invariant along the lines of "when there are no messages left to be delivered, the lender that the broker requested to commit, has committed". ProB quickly finds a counterexample to this assertion, involving a trace in which the key message is dropped by the network. This confirms what is already well known, that an unreliable and asynchronous network cannot be used to establish distributed agreement .

The traditional way to resolve this problem is to use an ACID transaction to commit the mortgage lender and insurance company when the user accepts their offer. We therefore modified our communication model so that the messages associated with offers were treated given top priority and never lost; in effect they were passed synchronously. ProB found that the invariant was now satisfied, helping to convince us that it was safe to use asynchronous message passing for the other communications.

One issue for a system such as the mortgage broker is the need for lenders and insurance companies to manage their commitments. It was decided to associate a deadline with each offer. Thus, though each party can send a message to cancel such an offer, this is not necessary. Indeed, when messages can get lost, it is not feasible to rely on them for cancellations, which are crucial for mortgage lenders to allow them to manage their reserve of funds. We were able to model in B, and implement in Java, this functionality (which we acknowledge introduces an additional requirement for synchronous communication into the underlying infrastructure).

### 3.3 Model-Based Trace-Checking

To provide us with traces, we wrote a generic "trace bean" for capturing key transitions in a common database. Trace points were easily identified as a result of the close correspondence between top-level model and implementation. By hand we added trace beans and tracing code at each such point. Thereafter, each execution of the system automatically added to our tracing database. It was now easy to extract traces using SQL queries, and a small amount of scripting code, to extract and transform traces into forms that can be model-checked by Spin or ProB [14]. As well as manual testing, we developed an automated "test-bot" to execute our implementation at random, and provide longer traces.

This technique proved quite successful. We found a number of errors, some of which were in our models, some of which were in the tracing code, and three of which were genuine (that is to say, it was the implementation that was wrong). Because we had not fully automated the approach, we were able to model-check only a subset of the traces that we captured. A potential problem with this approach is that it soon generates significant amounts of data. Trace-checking needs to be an efficient and automatic operation to cope with this.

As a result of using this technique, as well as the close team-working described above, we have good confidence that our top-level models, and particularly in the case of the mortgage broker, are consistent with our implementations. This particularly applies to the B/AMN model. We tended to use Promela/Spin for "throwaway" modelling, partly because it lacks rich data structures that can easily model complex databases. Also, most of our model evolution was through working directly with the B machine, so that our original UML models became gradually more and more out of date. Though U2B allows B to be generated from annotated UML, it does not support round-trip engineering – changes to the generated B are not transferred back to the original UML annotations.

It would also have been interesting to explore the use of model-based test-case generation. There is a recently released tool, BZTT [15], that provides this functionality for systems specified in the B or Z notations.

## 4. *Summary and Conclusions*

### 4.1 Lessons Learned
Based on our two case studies, both of which were relatively modest e-business applications developed by a small team, we can draw some tentative, but positive, conclusions.
- rapid modelling can be used to explore the consequences of design decisions, and provide early confidence in the chosen design. Where this approach is used, the model can be discarded once it has served its purpose, though
- there are benefits in retaining a model and keeping it consistent with the implementation during an evolutionary development,
- though formal review or inspection meetings can be used for this co-evolution, we found it more effective to use close team working, with all developers co-located in an open plan office, so

- developers and modellers can work together on implementation and models using pair "programming" techniques, and animation to validate models, and
- model-based testing, and in particular model-based trace-checking, is effective and this provides both a good reason and a further way to keep model and implementation consistent

Our main negative conclusion is that without good tool support, developers will not apply these techniques, or only reluctantly. This observation applies throughout the development cycle, from UML modelling to model-based testing. Some suggestions for enhanced tool support for formal models are given as future work, below.

## 4.2 Related Work

There is extensive literature on formal methods, though this typically assumes a top-down development methods, and the emphasis is often on formal verification of code, or design by stepwise refinement of a formal specification. Similarly, there is extensive literature on model-checking, though so far not much of this has been focussed on integrating this technology into the software development process.

AsmL [12] is a toolkit developed by Microsoft Research that integrates the Abstract State Machine approach into the debugging and error location phase of .NET software development. This is a technically impressive achievement, and it would be interesting to see tools for model-checking AsmL specifications, and for generating tests from them. The B and Z Testing Tools (BZTT [15]) are now available to use in generating tests from B machines and Z specifications; not surprisingly, there is industrial interest in these tools.

Model-driven development, or model-driven architecture [18], is a software method in an executable subset of UML is used to write platform independent models. These can be automatically converted into platform specific models by selecting an implementation target such as .NET or J2EE. There are a number of case studies that have successfully followed this approach.

In all the above approaches, a tool exists which support basically just one of the main tasks of development: design, implementation, or testing. For a notation to succeed, tool support needs to be provided across the whole spectrum of development activities. Model-checking and model-based testing are, I believe, crucial for a formal modelling approach to succeed. At present, many tools provide only point solutions, and some are only suitable for use in academia on small case studies. For formally-based development methods to be more widely adopted, there needs to be a set of tools which is mature, integrated, and supports a standard notation (or notations). What is encouraging is that, among the current diversity of provision, tools exist which demonstrate how formal models can be used to improve not only design, but also implementation and testing. If these tools could be easily be used together, their benefits would provide a powerful incentive to develop both model and implementation together.

## 4.3 Future Work: eXtreme Modelling

We can speculate on an future approach to software development, which we could call eXtreme Modelling, in which model and implementation are developed in parallel. In this approach, the model takes the place of, or supplements, the tests used in test-driven development [5]. This means the model is written just before it is implemented. Model-checkers can be used to confirm that the model satisfies desirable properties, expressed in temporal logic or as global system invariants, or to reject it if not. In practice, a range of models might be produced, each capturing a particular aspect of the system. These models can be used to generate tests, to insert assertions in the code and, in combination with a model-checker, to check execution traces. As the implementation evolves, so too does the model. Using pairs or teams of developers working closely together, models and code are kept in synchronisation. Moreover, as development progresses, the code is repeatedly tested against the model using the generated tests, inserted assertions, and model-based trace-checking. This automated regression testing provides confidence in the implementation, the model, and their mutual consistency.

At present we can only speculate on how effective or popular eXtreme Modelling would be if production quality tools were available and integrated into an integrated development environment. The main future work suggested here is to enhance tools that could support this approach:
- round-trip U2B to synchronise B machines and UML diagrams, to allow designers to edit generated models
- where executable subsets exist, round-trip code generation, to synchronise model and code, and to allow developers to edit generated code

- a suite of tools supporting animation, model-checking, automated insertion of assertions, model-based testing and trace-checking for models written in a single, standard notation.

Our experiences in developing our case studies suggest that co-evolution of model and implementation is indeed possible, at least in a University environment. Given enhanced tool support as indicated above industrial software developers would also, we believe, be happy to incorporate formal models in their evolutionary development processes.

This work forms part of the ABCD project, which is funded by the EPSRC (GR/M91013/01), whose support we gratefully acknowledge.

## *Bibliography*

[1] J.R. Abrial. *The B-Book*: *assigning programs to meanings*. Cambridge University Press, 1996.
[2] *Using the Extensible Model Checker XTL to Verify StAC Business Specifications.* Juan C. Augusto, Michael Leuschel, Michael Butler and Carla Ferreira. 3rd Workshop on Automated Verification of Critical Systems (AVoCS 2003), pp. 253--266, April 2003, Southampton (UK).
[3] Juan C. Augusto, Carla Ferreira, Andrew M. Gravell, Michael Leuschel, and Karen M.Y. NG, *The Benefits of Rapid Modelling for e-Business System Development,* accepted for publication at eCOMO'2003 4th Int. Workshop on Conceptual Modelling Approaches for e-Business, October 2003, Chicago.
[4] Kent Beck, eXtreme Programming explained, Addison-Wesley 2000.
[5] Kent Beck, Test-Driven Development, Addison-Wesley 2002.
[6] M.J. Butler, and C.F. Snook, *Verifying Dynamic Properties of UML Models by Translation to the B Language and Toolkit*, Proceedings UML 2000 Workshop.
[7] M.A. Leuschel and M.J. Butler, *ProB: A Model Checker for B*, FME 2003: Formal Methods, Springer-Verlag LNCS 2805, pages 855-874.
[8] E.M. Clarke, O. Grumberg, and D.A. Peled, *Model-Checking*, MIT Press 1999.
[9] Alistair Cockburn, *Agile Software Development*, Addison Wesley 2002.
[10] J. Dick and A. Faivre, *Automating the Generation and Sequencing of Test Cases from Model-Based Specifications*, FME 93, Springer Verlag 1993, pages 268-284.
[11] M.F. Doche, and A.M. Gravell. *Extraction of Abstraction Invariants for Data Refinement*. In Bert, Didier and Bowen, Jonathan P and Henson, Martin C and Robinson, Ken, Eds. *Proceedings ZB 2002*, pages 120-139, Grenoble.
[12] Y. Gurevich, W. Schulte, and M. Veanes, *Toward Industrial Strength Abstract State Machines*, Microsoft Research, MS-TR-2001-98, 2001.
[13] Gerard Holzmann. *The spin model checker*. IEEE Trans. on Software Engineering, 23(5):279–295, 1997.
[14] Howard M et al, *Model-Based Trace-Checking*, UK-Softest 2003, York, September 2003.
[15] B. Legeard, F. Peureux, and M. Utting, *Automated Boundary Testing from Z and B*, FME 2002, Springer Verlag LNCS 2391, pages 21-40.
[16] Nancy Lynch, *Distributed Algorithms*, Morgan Kaufmann 1996.
[17] S.J. Mellor, S. Tockey, R. Arthaud, and P. LeBlanc, *Software-platform-independent Precise Action Specifications for UML*, UML'98, pages 281-286.
[18] S.J. Mellor, K. Scott, A. Uhl, and D. Weise, *Model-Driven Architecture*, Springer LNCS 2426, 2002.
[19] W.W. Royce, *Managing the development of large software systems*, Proceedings IEEE WESCON, 1970, pages 1-9.
[20] W. Royce, *Software Project Management: a unified framework,* Addison Wesley 1998.
[21] James Rumbaugh, Ivar Jacobson and Grady Booch, *The Unified Modelling Language Reference Manual*, Addison Wesley 99.
[22] J.M. Spivey, *The Z Notation*, Prentice Hall 1992.
[23] J.M. Wing, *A Specifier's Introduction to Formal Methods*, Computer 23(9), 1990, pages 8-24.